\documentstyle[prl,aps,multicol,epsf]{revtex}
\draft 

\begin{document}

\title{Molecular motor that never steps backwards}

\author{Markus~Porto, Michael~Urbakh, and Joseph~Klafter}

\address{School~of~Chemistry, Tel~Aviv~University, 69978~Tel~Aviv, Israel}

\date{February 29, 2000}

\maketitle

\begin{abstract}
We investigate the dynamics of a classical particle in a one-dimensional
two-wave potential composed of two periodic potentials, that are
time-independent and of the same amplitude and periodicity. One of the periodic
potentials is externally driven and performs a translational motion with
respect to the other. It is shown that if one of the potentials is of the
ratchet type, translation of the potential in a given direction leads to motion
of the particle in the same direction, whereas translation in the opposite
direction leaves the particle localized at its original location. Moreover,
even if the translation is random, but still has a finite velocity, an
efficient directed transport of the particle occurs.
\end{abstract}

\vspace*{-3mm}
\vspace*{-2mm}

\begin{multicols}{2}
A particle subject to a spatially asymmetric but on large scale homogeneous
potential displays a symmetric diffusive motion, since the sole violation of
the $x \rightarrow -x$ symmetry is not sufficient to cause a net directional
transport. As already noted more than 100~years ago by Curie \cite{Curie:1894},
the additional breaking of time reversal $t \rightarrow -t$ symmetry [e.g.\ by
dissipation] may lead to a macroscopic net velocity, so that in this case
directed motion can result in the absence of any external force. Such systems,
known as thermal ratchets \cite{Feynman/Leighton/Sands:1963}, have been subject
of much activity, both theoretical
\cite{Ajdari/Prost:1993,Magnasco:1993,Astumian/Bier:1994,%
Prost/Chauwin/Peliti/Ajdari:1994,Millonas/Dykman:1994,%
Juelicher/Ajdari/Prost:1997,Astumian:1997,Dialynas/Lindenberg/Tsironis:1997,%
Thomas/Thornhill:1998,Qian:1998,Landa:1998,Sokolov:1998+99,%
Fisher/Kolomeisky:1999,Derenyi/Bier/Astumian:1999,Mateos:2000,%
Flach/Yevtushenko/Zolotaryuk:2000} and experimental
\cite{Rousselet/Salome/Ajdari/Prost:1994,%
Faucheux/Bourdieu/Kaplan/Libchaber:1995,Gorre/Ioannidis/Silberzan:1996,%
Linke/etal:1998+1999,Mennerat-Robilliard/etal:1999,%
Kettner/Reimann/Haenggi/Mueller:2000,Ajdari:2000}, partly motivated by possible
applicability to biological motors
\cite{Howard:1997,Huxley+Howard:1998,Okada/Hirokawa:1999}.

In this Letter we study the classical dynamics of a particle in a
one-dimensional two-wave potential. The total potential is composed of two
periodic potentials, that are time-independent and of equal amplitudes and
periodicities. One of the potentials is externally driven performing a
translational motion with respect to the other. It is shown that if, in
addition to the broken time reversal symmetry, the spatial symmetry is broken
for one of the potentials, the relative translation can result in a two-fold
behavior: {\it (i)}~Translation in one direction causes a deterministic motion
of the particle in the same direction, whereas {\it (ii)}~translation in the
opposite direction leaves the particle localized at its original location.
Thus, the total potential acts as a ratchet in the original sense. Moreover, an
efficient directed transport occurs even if the translation is random but still
has a finite velocity. The reason for the directed transport is the existance
of points of irreversibility in the particle trajectory. The high rate
transport stems from the fact that if the particle once gains a distance which
is an integer multiple of the potential period, this distance is preserved,
different from former ratchet systems driven by random fluctuations.

We consider a simple ratchet type potential $\Pi(x)$, which is assumed to be
continuous but not necessarily differentiable. It has a periodicity $b$, so
that $\Pi(x+b) = \Pi(x)$ $\forall x$, an amplitude $\Pi_0 = \max \Pi(x) = -\min
\Pi(x)$, and one minimum is located at $x = 0$, i.e.\ $\Pi(0) = -\Pi_0$. We
also assume that the potential $\Pi(x)$ has only one minimum and one maximum
per period $b$, so that $\partial \Pi(x)/\partial x$ changes sign only twice
for $x \in [0, b)$, cf.\ \cite{Note1}. The total potential $V(x,\gamma)$ is
composed of two, not necessarily identical potentials $\Pi(x)$ and $\Pi'(x)$,
i.e.\ $V(x,\gamma) \equiv \Pi(x) + \Pi'(x-\gamma)$, where $\gamma$ defines the
translation. Because of the periodicity of the potentials $\Pi(x)$, the
potential $V(x,\gamma)$ is periodic in both arguments, so that $V(x,\gamma+b) =
V(x,\gamma)$ $\forall \gamma$. In this potential landscape, the deterministic
equation of motion of a particle of mass $m$ reads as
\begin{equation}\label{eq:motion}
m \ddot{x} + \eta \dot{x} + \frac{\partial V(x,\gamma)}{\partial x} = 0 \quad,
\end{equation}
where the damping is denoted by $\eta$. It should be emphasized that, through
the translation, energy is being permanently fed into the system, and that this
energy has to be dissipated, i.e.\ $\eta > 0$. Otherwise the particle will gain
energy until it decouples from the potential.

First, we restrict ourselves to the case characterized by an overdamped motion
with $\eta/[(2 \pi/b) \; \sqrt{m \Pi_0}] \gg 1$ and by a slow translation with
$|\dot{\gamma}|/[2 \pi \sqrt{\Phi_0/m}] \ll 1$, where $\dot{\gamma}$ is the
translation velocity, and relax these restrictions towards the end. In this
limit, as the translation $\gamma$ is varied, the particle either {\it
(i)}~moves slowly remaining at the local potential minimum or {\it (ii)}~if the
minimum ceases to exists, it jumps to the next minimum following the potential
slope. The latter happens instantaneously on the time scale of the translation.
In order to obtain the observables, which are the trajectory $x$ and the
average velocity $\overline{\dot{x}}$, it is therefore sufficient to study the
behavior of the total potential $V(x,\gamma)$ and evaluate the positions of the
minima. Below this limit is refered as `quasistatic'. The initial conditions at
$t = 0$ are chosen as $x = 0$ and $\gamma = 0$, so that the particle is located
at a potential minimum. In what follows, we use the abbreviations $\tilde{x}
\equiv x/b \bmod 1$ and $\tilde{\gamma} = \gamma/b \bmod 1$.

For simplicity, we start the discussion with a particular example for the
potential $\Pi(x)$,
\begin{equation}\label{eq:ratchet}
\Pi_{\xi}(x) \equiv \Pi_0 \left\{
\begin{array}{ll}
\displaystyle -1 + 2 \frac{\tilde{x}}{\xi} &
\mbox{if $\displaystyle \tilde{x} \le \xi$}\\[5mm]
\displaystyle 1 - 2 \frac{\tilde{x}- \xi}{1 - \xi} &
\mbox{if $\displaystyle \tilde{x} > \xi$}
\end{array}\right. \quad,
\end{equation}
which is piecewise linear, although the arguments below
\linebreak\hspace*{1mm}\par\vspace*{-2\baselineskip}\hspace*{1mm}\par
\begin{figure}[h]
\vspace*{-3mm}
\def\epsfsize#1#2{0.55#1}
\noindent\hfill\epsfbox{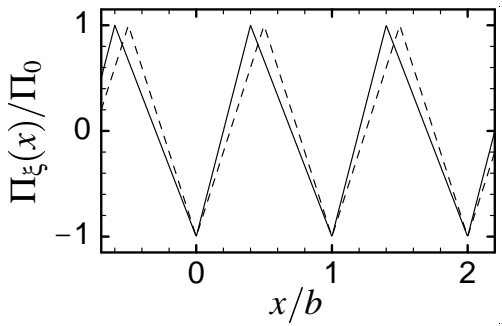}\hfill%
\vspace*{1.5mm}
\caption{Plot of the ratchet given by eq.~(\ref{eq:ratchet}) with
asymmetry parameters $\xi = 2/5$ [solid line] and $\xi' = 1/2$ [dashed
line, symmetric case].}
\label{figure:ratchet}
\vspace*{-2mm}
\end{figure}

\noindent
apply analogously for
other ratchet type potentials as well. The parameter $\xi \in (0, 1 )$
determines the asymmetry of the potential, with $\xi = 1/2$ being the symmetric
case. In Fig.~\ref{figure:ratchet} two realizations are shown for $\xi = 2/5$
and $\xi' = 1/2$. In order to introduce only one asymmetry, we choose the
translated potential $\Pi'_{\xi'}(x)$ to be symmetric, i.e.\ $\xi' = 1/2$. For
the other potential $\Pi_{\xi}(x)$ we use $\xi < 1/2$, since the case $\xi >
1/2$ can be mapped on the case $\xi < 1/2$ by replacing the asymmetry $\xi$ by
$1-\xi$ and the translation $\gamma$ by $-\gamma$. The case $\xi = \xi' = 1/2$,
where both potentials are symmetric, is excluded, since then the total
potential $V_{\xi,\xi'}(x,\gamma)$ becomes piecewise flat, and the
`quasistatic' treatment of eq.~(\ref{eq:motion}) is no longer valid
\cite{Note2}.
\linebreak\hspace*{1mm}\par\vspace*{-2\baselineskip}\hspace*{1mm}\par

\begin{figure}[h]
\vspace*{-3mm}
\def\epsfsize#1#2{0.39#1}
\noindent\hfill\epsfbox{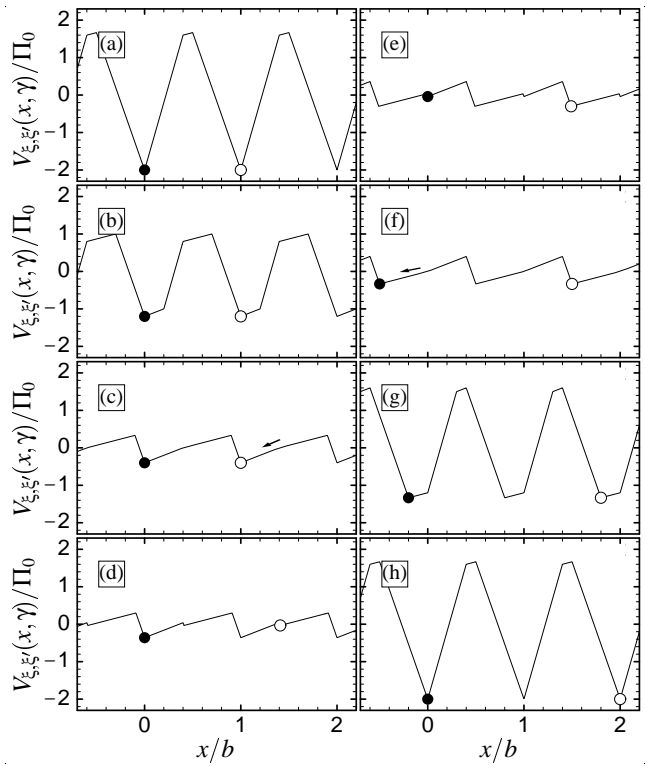}\hfill%
\vspace*{1.5mm}
\caption{Time evolution of the total potential $V_{\xi,\xi'}(x,\gamma)$
for the ratchet given by eq.~(\ref{eq:ratchet}). In parallel, the
respective positions of the particle are shown, both for $\dot{\gamma} >
0$ [full circles, the time evolves from (a) to (h)] and $\dot{\gamma} <
0$ [open circles, the time evolves in the opposite direction from (h) to
(a)]. The parameters are $\xi = 2/5$ and $\xi' = 1/2$, and snapshoots are
taken at (a)~$\tilde{\gamma} = 0$, (b)~$\tilde{\gamma} = 1/5$,
(c)~$\tilde{\gamma} = 2/5$, (d)~$\tilde{\gamma} = 41/100$,
(e)~$\tilde{\gamma} = 49/100$, (f)~$\tilde{\gamma} = 1/2$,
(g)~$\tilde{\gamma} = 4/5$, and (h)~$\tilde{\gamma} = 0$. The arrows
indicate the direction of irreversible motion of the particle which
occurs between snapshots (e) and (f) [full circle] and (d) and (c) [open
circle].}
\label{figure:motion}
\vspace*{-2mm}
\end{figure}

Let us first discuss a translation with a constant translation velocity
$\dot{\gamma} = {\rm const}$, so that $\gamma = \dot{\gamma} t$, where the
velocity can be either $\dot{\gamma} < 0$ or $\dot{\gamma} > 0$. In
Fig.~\ref{figure:motion} shown is one cycle of a translation by $-b$
[$\dot{\gamma} < 0$] and $b$ [$\dot{\gamma} > 0$] for an example with $\xi=
2/5$ and $\xi'= 1/2$. In the case $\dot{\gamma} < 0$ [the open circles in
Fig.~\ref{figure:motion}, the time evolves from (h) to (a)] the particle moves
a distance $-b$, whereas in the case $\dot{\gamma} > 0$ [the full circles in
Fig.~\ref{figure:motion}, the time evolves from (a) to (h)] the particle,
although moving locally, returns to its starting point. After that, the whole
cycle starts over again. Hence, the particle moves either with average velocity
$\overline{\dot{x}} = \dot{\gamma}$ for $\dot{\gamma} < 0$ or
$\overline{\dot{x}} = 0$ for $\dot{\gamma} > 0$. For the opposite asymmetry
$\xi > 1/2$, the particle remains at $x = 0$ for $\dot{\gamma} < 0$ and moves
with average velocity $\overline{\dot{x}} = \dot{\gamma} > 0$ in the case of an
opposite translation.

\begin{figure}[h]
\vspace*{-3mm}
\def\epsfsize#1#2{0.55#1}
\noindent\hfill\epsfbox{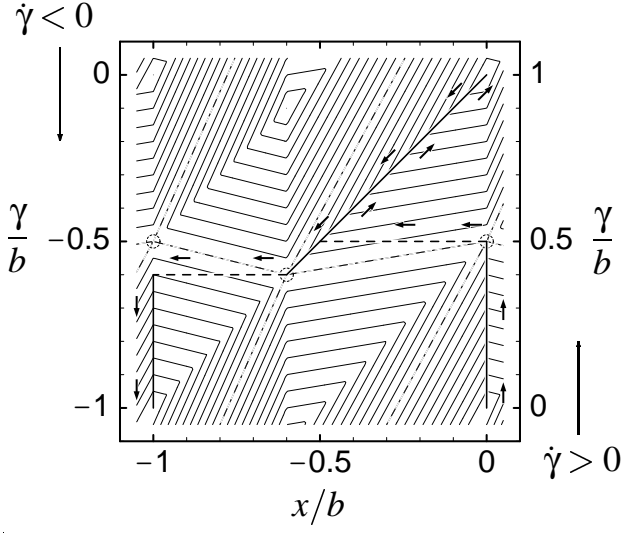}\hfill%
\vspace*{1.5mm}
\caption{Contour plot of the total potential $V_{\xi,\xi'}(x,\gamma)$ for
the ratchet given by eq.~(\ref{eq:ratchet}) with $\xi = 2/5$ and $\xi' =
1/2$, the solid equipotential lines are placed at $V_{\xi,\xi'}(x,\gamma)
= \pm n \Pi_0/5$ with $1 \le n \le 10$ integer, and the dashed-dotted
equipotential lines indicate $V_{\xi,\xi'}(x,\gamma) = 0$. The respective
trajectories of a particle starting at position $x = 0$ at translation
$\gamma = 0$ are shown for both $\dot{\gamma} < 0$ [left side] and
$\dot{\gamma} > 0$ [right side] with thick lines, which are solid for the
part of the trajectory where the particle remains in the minimum, and
dashed for the irreversible jumps. The arrows indicate the time
development for $\dot{\gamma} < 0$ [downward arrows] and $\dot{\gamma} >
0$ [upward arrows]. The points of irreversibility ${\cal I} = \{
(\tilde{x}, \tilde{\gamma}) | (\tilde{x}, \tilde{\gamma}) = ( 0, 1/2), (
2/5, 2/5) \}$ are marked by dashed circles.}
\label{figure:globaltopology}
\vspace*{-2mm}
\end{figure}

This behavior can be generally understood from the properties of the total
potential $V(x,\gamma)$. In Fig.~\ref{figure:globaltopology} shown are the
potential $V(x,\gamma)$ and, for both $\dot{\gamma} < 0$ [left side] and
$\dot{\gamma} > 0$ [right side], the particle position $x$ for a particle
located at $x = 0$ for translation $\gamma = 0$, as it follows the changes of
the potential $V(x,\gamma)$. Due to the asymmetry of one of the constituting
potentials, there are points of instability ${\cal I}$ in the $(\tilde{x},
\tilde{\gamma})$ plane, where a local minimum of the potential $V(x,\gamma)$
ceases to exist. Hence, when the minimum disappears, a particle located at such
a point performs an irreversible motion jumping to the next minimum, see
Figs.~\ref{figure:motion}(c),(f) and Fig.~\ref{figure:globaltopology}. In the
example considered here, this new minimum moves in the direction of the jump
and a net transport occurs in the case $\dot{\gamma} < 0$, whereas in the case
$\dot{\gamma} > 0$ it moves in the opposite direction and cancels the distance
\linebreak\hspace*{1mm}\par\vspace*{-2\baselineskip}\hspace*{1mm}\par

\noindent
gained by the jump. Under our restrictions on $\Pi(x)$, in particular due to
the equality of the potential amplitudes, all points $(\tilde{x},
\tilde{\gamma}) \in {\cal I}$ satisfy $V(x,\gamma) = 0$. Let ${\cal C} \equiv
\{ (\tilde{x}, \tilde{\gamma}) | V(x,\gamma) = 0 \} \supset {\cal I}$ be the
set of all pairs $(\tilde{x}, \tilde{\gamma})$ for which the potential
$V(x,\gamma) = 0$. Since the potentials $\Pi(x)$ are continuous, the topology
of ${\cal C}$ is such that it consists of connected points, that form paths and
intersections, see Fig.~\ref{figure:globaltopology}. The intersections
correspond to the points of irreversibility ${\cal I}$ in which a minimum of
$V(x,\gamma)$ with respect to $x$ ceases to exist. For the particular choice of
the potentials $\Pi(x)$ given by eq.~(\ref{eq:ratchet}) one obtains the two
points ${\cal I}_{\xi,\xi'} = \{ (\tilde{x}, \tilde{\gamma}) | (\tilde{x},
\tilde{\gamma}) = (0, 1-\xi'), (\xi, \xi) \}$. Within the `quasistatic' limit,
jumps in the direction given by the potentials' asymmetry occur if the particle
reaches these points \cite{Note3}.

\begin{figure}[h]
\vspace*{-3mm}
\def\epsfsize#1#2{0.47#1}
\noindent\hfill\epsfbox{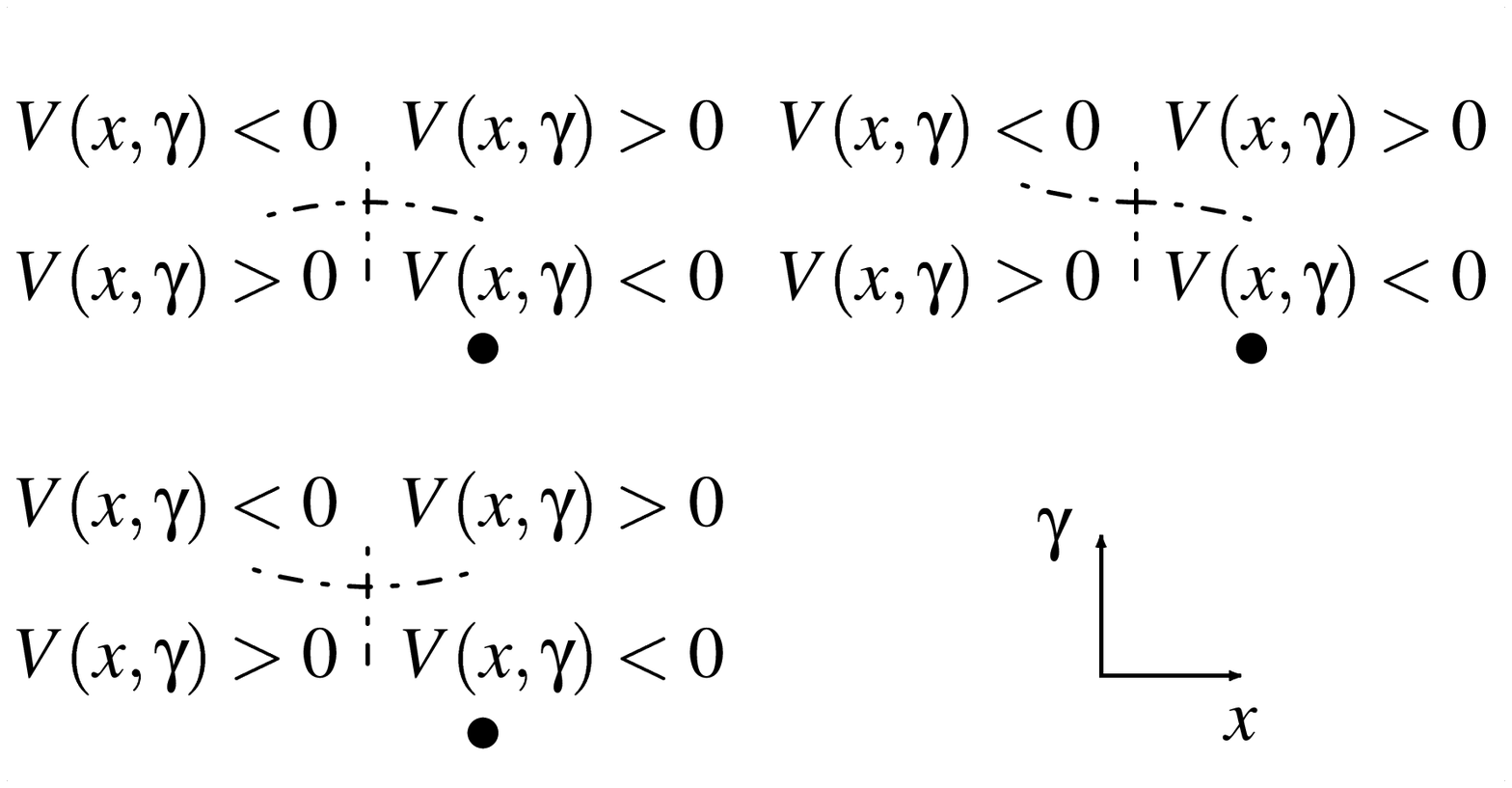}\hfill%
\vspace*{1.5mm}
\caption{Possible topologies for the set ${\cal C}$ around the points of
irreversibility ${\cal I}$. The dashed lines indicate $V(x,\gamma) = 0$,
and the areas with $V(x,\gamma) > 0$ and $V(x,\gamma) < 0$ are indicated.
The particle's position is shown as full circle.}
\label{figure:localtopology}
\vspace*{-2mm}
\end{figure}

In order to decide in general, if for a given choice of potentials $\Pi(x)$ a
directed transport is possible within the `quasistatic' limit, one has to
examine the set ${\cal C}$ around the intersection points ${\cal I}$. In
Fig.~\ref{figure:localtopology} the possible scenarios are shown, each with
sketched `horizontal' and `vertical' lines corresponding to $V(x,\gamma) = 0$
around the intersection points. Concerning the question of irreversible jumps,
namely if the particle reaches
\linebreak\hspace*{1mm}\par\vspace*{-2\baselineskip}\hspace*{1mm}\par

\noindent
the points of irreversibility as the translation
is monotonously varied, one has to examine the behavior of the horizontal line,
i.e.\ how this line is bent at the intersection point with respect to the
direction of the translation. In the upper left part of
Fig.~\ref{figure:localtopology} the horizontal line is bent downwards towards
smaller values of $\gamma$ and hence opposite to the direction of the
translation $\dot{\gamma} > 0$ on both sides of the intersection point. This
means that {\it (i)}~the minimum in which the sketched particle is located
moves towards the intersection point as $\gamma$ is increased [because of the
downward bending on the particle's side], and {\it (ii)}~the minimum ceases to
exist at the intersection point with a local slope such that the particle
performs a jump leftwards to smaller values of $x$ [because of the downward
bending on the side opposite to the paricle]. In the lower left part of
Fig.~\ref{figure:localtopology} the horizontal line is bent upwards on the
particle's side of the intersection point, so that the minimum does {\it not}
move towards the intersection as $\gamma$ is increased, but remains always
right to it at larger values of $x$. This behavior is independent of the
bending on the side of the intersection point opposite to the particle, either
upward, as shown in the figure, or downward. In the upper right part of
Fig.~\ref{figure:localtopology} a third theoretically possible topology is
shown. In this case, the particle reaches the intersection point [because of
the downward bending on the particle's side], but the intersection point is
{\it not} a point where the
\linebreak\hspace*{1mm}\par\vspace*{-2\baselineskip}\hspace*{1mm}\par

\noindent
minimum ceases to exist [because of the upward
bending on the side opposite to the particle]. However, this topology cannot
occur under our restrictions on the potentials $\Pi(x)$. For the potentials
$\Pi(x)$ given by eq.~(\ref{eq:ratchet}) and for $\xi \not= 1/2$ or $\xi' \not=
1/2$ the resulting topology of $\cal C$ is always such that the points of
irreversibility are reached within the `quasistatic' limit. Hence, depending on
the asymmetry of the potentials, one observes transport for either
$\dot{\gamma} < 0$ or $\dot{\gamma} > 0$, and no transport in the case of an
opposite translation.

We can extend the model beyond the translation with a constant velocity, by
assuming for instance another simple scenario with an oscillatory translation
of the form $\gamma = \gamma_0 + \gamma_1 \sin(2 \pi \omega t)$ with a driving
frequency $\omega$ and $\eta |\gamma_1| \omega b/\Pi_0 \ll 1$. In this case,
one finds an average velocity $\overline{\dot{x}} = \pm n b \omega$ with $n \ge
0$ integer, where the sign depends on the asymmetry of the potentials $\Pi(x)$.
The actual value of $n$ depends on the offset $\gamma_0$ and on the amplitude
$\gamma_1$, since these values determine how many times the points of
irreversibility ${\cal I}$ are reached during one cycle. For the potentials
$\Pi(x)$ given by eq.~(\ref{eq:ratchet}) with $\xi = 2/5$ and $\xi' = 1/2$, a
choice of $\gamma_0 = 45 b/100$ and $\gamma_1 = b/20$ results in an average
velocity of $\overline{\dot{x}} = -b \omega$. This means that in order to make
the particle gain a distance $b$, the translation has only to be varied twice
by $b/10$.

\begin{figure}[h]
\vspace*{-3mm}
\def\epsfsize#1#2{0.56#1}
\noindent\hfill\epsfbox{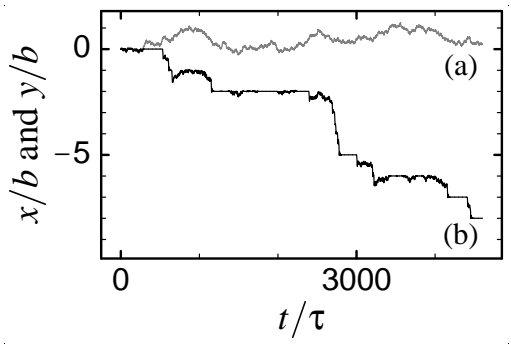}\hfill%
\vspace*{1.5mm}
\caption{Plot of a random walk trace $y$ [(a), in grey] used as
translation and the numerically obtained particle trajectory $x$ [(b), in
black] for asymmetry parameters $\xi = 2/5$ and $\xi' = 1/2$, and
dissipation constant $\eta/[(2 \pi/b) \sqrt{m \Phi_0}] = 0.2$. The random
walker moves with velocity $|\dot{y}|/(2 \pi \sqrt{\Phi_0/m}) = 0.02$ and
chooses the direction anew every $t/\tau = 1$ with $\tau \equiv [(2
\pi/b) \sqrt{\Phi_0/m}]^{-1}$, thus gaining a distance $|\Delta y|/b =
0.02$ during each step.}
\label{figure:randomwalk}
\vspace*{-2mm}
\end{figure}

As a third possible scenario for the translation $\gamma$, we assume $\gamma$
to vary randomly by following the trace of a random walker $y$ that has locally
a finite constant velocity $0 < |\dot{y}| = {\rm const}$, but an average
velocity $\overline{\dot{y}}$ = 0. For this scenario we relax the restriction
of overdamped motion to demonstrate the validity and accuracy of the
`quasistatic' treatment, and numerically integrate eq.~(\ref{eq:motion}) with a
finite damping and a finite translation velocity. However, one has to keep the
restriction $|\dot{\gamma}|/[2 \pi \sqrt{\Phi_0/m}] \ll 1$, since otherwise the
changes in the potential due to the translation are too fast for the particle
to follow, and the particle decouples from the potential. In
Fig.~\ref{figure:randomwalk} a realization of a random walker trace $y$ and the
resulting particle trajectory $x$ for $\xi = 2/5$ and $\xi' = 1/2$ are shown.
We would like to emphasize that, despite the relatively small damping, the
numerically obtained trajectory differs from the one obtained under overdamped
conditions only by small oscillations. Although the random walker, and hence
the
\linebreak\hspace*{1mm}\par\vspace*{-2\baselineskip}\hspace*{1mm}\par

\noindent
translation, have an average velocity $\overline{\dot{y}} = 0$ resp.\
$\overline{\dot{\gamma}} = 0$, the particle's average velocity
$\overline{\dot{x}}$ is nonzero. The sign of the velocity depends on the
asymmetry of the potentials $\Pi(x)$, for our choice of asymmetry one finds
$\overline{\dot{x}} < 0$ as expected. The transport is much more efficient than
in former ratchet systems driven by random fluctuations. Here, if the particle
gains a distance $d(t) \equiv |x(t)|$ equal to an integer multiple of the
potential period $b$ at a time $t_n$, $d(t_n) = n b$, the distance will never
become smaller again, and $d(t) \ge n b$ $\forall t \ge t_n$. Putting this in
the context of molecular motors, it means that the suggested molecular motor
never executes a step backwards. These backward steps usually limit the
efficiency of the motor \cite{Juelicher/Ajdari/Prost:1997}.
\linebreak\hspace*{1mm}\par\vspace*{-2\baselineskip}\hspace*{1mm}\par

Different realizations of the model can be thought of. One possibility is to
put a small particle in a potential created by an optical tweezer such as in
\cite{Faucheux/Bourdieu/Kaplan/Libchaber:1995}, but with two time-independent
potentials added on top of each other at the same place and having a certain
phase shift. Although the phase shift is changed randomly, but with a finite
velocity, an efficient transport of the particle is predicted in a certain
direction determined by the asymmetry of one of the potentials. It is possible,
for instance, to induce the translation of the potential through the coupling
of a varying internal degree of freedom of the particle to the non-translated
potential \cite{Porto/Urbakh/Klafter:2000}. This opens various possibilities
for the construction of micro- and nanoscale devices such as pumps and motors
based on the presented ratchet type system.

Financial support from the Israel Science Foundation, the German Israeli
Foundation, and DIP and SISI\-TOMAS grants is gratefully acknowledged. M.P.\
gratefully acknowledges the Alexander von Humboldt Foundation (Feodor Lynen
program) for financial support.

\end{multicols}


\vspace*{-7mm}
\begin{thebibliography}{99}
\vspace*{-18mm}

\bibitem{Curie:1894}
 M.P.~Curie,
 J.\ Phys.\ (Paris)~III~{\bf 3}, 393 (1894).

\bibitem{Feynman/Leighton/Sands:1963}
 R.P.~Feynman, R.B.~Leighton, and M.~Sands,
 {\it The Feynman Lectures on Physics}
 (Addison-Wesley, Reading, 1963),
 Vol.~1, Chap.~46.

\bibitem{Ajdari/Prost:1993}
 A.~Ajdari and J.~Prost,
 C.R.\ Acad.\ Sci.\ Paris~II~{\bf 315}, 1635 (1993).

\bibitem{Magnasco:1993}
 M.O.~Magnasco,
 Phys.\ Rev.\ Lett.~{\bf 71}, 1477 (1993).

\bibitem{Astumian/Bier:1994}
 R.D.~Astumian and M.~Bier,
 Phys.\ Rev.\ Lett.~{\bf 72}, 1766 (1994).

\bibitem{Prost/Chauwin/Peliti/Ajdari:1994}
 J.~Prost, J.-F.~Chauwin, L.~Peliti, and A.~Ajdari,
 Phys.\ Rev.\ Lett.~{\bf 72}, 2652 (1994).

\bibitem{Millonas/Dykman:1994}
 M.M.~Millonas and M.I.~Dykman,
 Phys.\ Lett.~A~{\bf 185}, 65 (1994).

\bibitem{Juelicher/Ajdari/Prost:1997}
 F.~J\"ulicher, A.~Ajdari, and J.~Prost,
 Rev.\ Mod.\ Phys.~{\bf 69}, 1269 (1997).

\bibitem{Astumian:1997}
 R.D.~Astumian,
 Science~{\bf 276}, 917 (1997).

\bibitem{Dialynas/Lindenberg/Tsironis:1997}
 T.E.~Dialynas, K.~Lindenberg, and G.P.~Tsironis,
 Phys.\ Rev.~E~{\bf 56}, 3976 (1997).

\bibitem{Thomas/Thornhill:1998}
 N.~Thomas and R.A.~Thornhill,
 J.\ Phys.~D~{\bf 31}, 253 (1998).

\bibitem{Qian:1998}
 H.~Qian,
 Phys.\ Rev.\ Lett.~{\bf 81}, 3063 (1998).

\bibitem{Landa:1998}
 P.S.~Landa,
 Phys.\ Rev.~E~{\bf 58}, 1325 (1998).

\bibitem{Sokolov:1998+99}
 I.M.~Sokolov,
 Europhys.\ Lett.~{\bf 44}, 278 (1998);
 J.\ Phys.~A~{\bf 32}, 2541 (1999);
 Phys.\ Rev.~E~{\bf 60}, 4946 (1999).

\bibitem{Fisher/Kolomeisky:1999}
 M.E.~Fisher and A.B.~Kolomeisky,
 Proc.\ Natl.\ Acad.\ Sci.\ USA~{\bf 96}, 6597 (1999).

\bibitem{Derenyi/Bier/Astumian:1999}
 I.~Der\'enyi, M.~Bier, and R.D.~Astumian,
 Phys.\ Rev.\ Lett.~{\bf 83}, 903 (1999).

\bibitem{Mateos:2000}
 J.L.~Mateos,
 Phys.\ Rev.\ Lett.~{\bf 84}, 258 (2000).

\bibitem{Flach/Yevtushenko/Zolotaryuk:2000}
 S.~Flach, O.~Yevtushenko, Y.~Zolotaryuk,
 Phys.\ Rev.\ Lett.~{\bf 84}, 2358 (2000).

\bibitem{Rousselet/Salome/Ajdari/Prost:1994}
 J.~Rousselet, L.~Salome, A.~Ajdari, and J.~Prost,
 Nature~{\bf 370}, 446 (1994).

\bibitem{Faucheux/Bourdieu/Kaplan/Libchaber:1995}
 L.P.~Faucheux, L.S.~Bourdieu, P.D.~Kaplan, and A.J.~Libchaber,
 Phys.\ Rev.\ Lett.~{\bf 74}, 1504 (1995).

\bibitem{Gorre/Ioannidis/Silberzan:1996}
 L.~Gorre, E.~Ioannidis, and P.~Silberzan,
 Europhys.\ Lett.~{\bf 33}, 267 (1996).

\bibitem{Linke/etal:1998+1999}
 H.~Linke, W.~Sheng, A.~Lofgren, H.Q.~Xu, P.~Omling, P.E.~Lindelof,
 Europhys.\ Lett.~{\bf 44}, 343 (1998); {\it ibid}~{\bf 45}, 406(E) (1999).

\bibitem{Mennerat-Robilliard/etal:1999}
 C.~Mennerat-Robilliard, D.~Lucas, S.~Guibal, J.~Tabosa, C.~Jurczak,
 J.-Y.~Courtois, and G.~Grynberg,
 Phys.\ Rev.\ Lett.~{\bf 82}, 851 (1999).

\bibitem{Kettner/Reimann/Haenggi/Mueller:2000}
 C.~Kettner, P.~Reimann, P.~H\"anggi, and F.~M\"uller,
 Phys.\ Rev.~E~{\bf 61}, 312 (2000).

\bibitem{Ajdari:2000}
 A.~Ajdari,
 Phys.\ Rev.~E~{\bf 61}, R45 (2000).

\bibitem{Howard:1997}
 J.~Howard,
 Nature~{\bf 389}, 561 (1997).

\bibitem{Huxley+Howard:1998}
 A.~Huxley,
 Nature~{\bf 391}, 239 (1998);
 J.~Howard,
 Nature~{\bf 391}, 240 (1998).

\bibitem{Okada/Hirokawa:1999}
 Y.~Okada and N.~Hirokawa,
 Science~{\bf 283}, 1152 (1999).

\bibitem{Note1}
 Since we do not require the potential $\Pi(x)$ to be differentiable, we do not
 assume the derivation to have two roots $\partial \Pi(x)/\partial x = 0$, but
 two sign changes.

\bibitem{Note2}
 If both potentials $\Pi(x)$ and $\Pi'(x)$ are spatially symmetric the total
 potential $V(x,\gamma)$ looses its ratchet character. Due to the unbroken
 spatial symmetry, either no transport occurs for both $\dot{\gamma} < 0$ and
 $\dot{\gamma} > 0$, or symmetric transport is observed in both directions,
 $\dot{x} < 0$ for $\dot{\gamma} < 0$ and $\dot{x} > 0$ for $\dot{\gamma} > 0$.

\bibitem{Note3}
 In general, both points of irreversibility can contribute to a directed
 transport, depending on the further evolution of the translation. For the
 example considered with $\xi = 2/5$ and $\xi' = 1/2$, if one translates the
 second potential from $\tilde{\gamma} = 0$ to $\tilde{\gamma} = 1/2$ and then
 back to $\tilde{\gamma} = 0$, a first jump occurs when reaching
 $\tilde{\gamma} = 1/2$ with $\dot{\gamma} > 0$, and a second jump occurs when
 passing $\tilde{\gamma} = 2/5$ for the second time with $\dot{\gamma} < 0$,
 both contributing to the directed transport.

\bibitem{Porto/Urbakh/Klafter:2000}
 M.~Porto, M.~Urbakh, and J.~Klafter
 (unpublished).

\end{thebibliography}
\end{document}